# Multi-Transfer Learning Techniques for Detecting Auditory Brainstem Response

Fatih Özyurt[1], Jafar Majidpour[2], Tarik A. Rashid[3], Amir Majidpour[4], Canan Koç[1]

[1]Department of Software Engineering, Faculty of Engineering, Firat University, Elazig, Turkey

[2]Department of Computer Science, University of Raparin, Rania, Iraq.

[3]Computer Science and Engineering Department, University of Kurdistan Hewlêr, Erbil, Iraq.

[4]Audiology Department, School of Rehabilitation, Shahid Beheshti University of Medical Sciences, Tehran, Iran

## Abstract

The assessment of the well-being of the peripheral auditory nerve system in individuals experiencing hearing impairment is conducted through auditory brainstem response (ABR) testing. Audiologists assess and document the results of the ABR test. They interpret the findings and assign labels to them using reference-based markers like peak latency, waveform morphology, amplitude, and other relevant factors. Inaccurate assessment of ABR tests may lead to incorrect judgments regarding the integrity of the auditory nerve system; therefore, proper Hearing Loss (HL) diagnosis and analysis are essential. To identify and assess ABR automation while decreasing the possibility of human error, machine learning methods, notably deep learning, may be an appropriate option. To address these issues, this study proposed deep-learning models using the transfer-learning (TL) approach to extract features from ABR testing and diagnose HL using support vector machines (SVM). Pre-trained convolutional neural network (CNN) architectures like AlexNet, DenseNet, GoogleNet, InceptionResNetV2, InceptionV3, MobileNetV2, NASNetMobile, ResNet18, ResNet50, ResNet101, ShuffleNet, and SqueezeNet are used to extract features from the collected ABR reported images dataset in the proposed model. It has been decided to use six measures accuracy, precision, recall, geometric mean (GM), standard deviation (SD), and area under the ROC curve to measure the effectiveness of the proposed model. According to experimental findings, the ShuffleNet and ResNet50 models' TL is effective for ABR to diagnose HL using an SVM classifier, with a high accuracy rate of 95% when using the 5-fold cross-validation method.

## 1. Introduction

### 1.1. Background

Audiologists usually use behavioral auditory tests to evaluate auditory sensitivity, but these subjective methods aren’t suitable for all groups, especially for individuals without the ability to respond, like infants, adults with learning disorders, and those suspected of non-organic hearing loss [1-3]. Audiologists, to compensate for this limitation, use objective tests such as auditory brain stem response (ABR) [4, 5]. ABRs are short-latency potentials evoked from the brain stem in response to acoustic stimulus [6] and can be recorded in a duration of less than 10-15 milliseconds (ms) after stimulus onset [2]. ABR is typically measured in an anechoic audiometric room using three surface electrodes on the scalp [5].

After preparing a patient for the test and performing the ABR test, an expert and trained audiologist would analyze and interpret ABR waveforms using the latency, inter-peak latency, and amplitude of ABR peaks [7]. As interpreting the ABR results needs lots of Specialty and skills, less expert clinicians may misinterpret the results and misdiagnose [8]. Additionally, the presence of distorted ABR waveform morphology in some specific disorders such as Central Auditory Processing Disorders (CAPD) further raises the risk of misdiagnosis [2]. So, it needs to automate ABR analysis to assist audiologists in making correct judgments and reducing hearing loss diagnosis errors.

In light of these findings, we propose a novel ABR detection methodology that uses ML and DL tools. The following is a list of the contributions made by the suggested approach.

- Analysing the efficacy of TL as a foundational approach to the ABR problem.
- Resolve the problem of deep learning networks' time consumption.

### 1.2. Motivation

Artificial intelligence (AI) refers to the use of computers to automate complex tasks generally performed by humans [8, 9]. Machine learning (ML) is a type of AI that is a powerful method of data analysis that is based on the concepts of learning and discovering data patterns rather than being programmed [10]. The use of ML in ABR is a rapidly developing field as an objective means to automate the analysis of the ABR waveforms and to achieve a better HL diagnosis, reduce clinical workload, and save more time for performing more ABR tests [6]. A machine learning technology called deep learning teaches computers to carry out tasks that people accomplish naturally. Deep learning allows a computer model to learn the information necessary to categorize data, such as photos, text, or audio. Deep learning models can occasionally perform more accurately than people. The models are trained using multi-layered neural network architectures and a sizable tagged dataset. Deep learning models are particularly good at identifying difficult-to-identify inputs and detecting ABRs because they may learn by focusing on their commonalities [11-13]. In the field of deep learning, [14] used 614 subjects with an average age between 18 and 90 years, 348 men and 266 females, using a long short-term memory (LSTM) model rich to 92.91% accuracy. With a different strategy, [15] offered a technique based on spectral feature extraction that would speed up detection without compromising accuracy. The technique uses a built-in feature-frequency vector that is supplied to an artificial neural network (ANN), and it produces the highest-performing model with sensitivity, specificity, and accuracy values of 87%, 89%, and 88%, respectively.

Since the peaks in the ABR have a very small amplitude, it is necessary to repeat the stimuli a large number of times [16]. The number of repetitions of stimuli in the test is usually 2000, which takes about 3 minutes to record [16, 17]. Considering that repeatability is necessary to determine the peaks and check the presence or absence of the peaks if the test is repeated twice at a certain intensity level, 6 minutes are required to record the waveform at that specific intensity [17]. All these estimates are provided when the audiologist performing the test is skilled enough to correctly determine and interpret the peaks, because if the audiologist is less experienced, more repetitions are needed to determine the peaks, and the possibility of misdiagnosis would also increase. Because of the reported facts, our main objective was to develop a machine learning algorithm to improve the automatic classification of recorded auditory brainstem responses as normal or pathological.

Some studies have used linear discriminant analysis (LDA) as an ABR classifier, but Sadeghian et al. [18] showed that the speech ABR of five English vowels can be classified with an impressive 83.33% accuracy using sustained response features and 38.33% accuracy using transient response features. LDA was used to characterize speech-evoked auditory brainstem responses (ABR) of three CV phones (/ba/, /da/, and /ga/) and three musical notes, as described in Losorelli et al.'s published work [19], which employed the same classifier. The authors reported accuracy scores ranging from 62.8% to 75%.

Some of these technological methods for ABR analysis incorporated the use of feature extraction methods to retrieve pertinent data from ABR evaluations and machine learning (ML) to categorize the outcomes. Dobrowolski et al. propose a classification method for ABR using wavelet decomposition and a Support Vector Machine (SVM) network. The approach demonstrates promising results in accurately categorizing ABR signals (97%), with a sample size of 130 participants [17]. A frequency-following response (FFR) using Chinese lexical tones as stimuli was shown to have an accuracy range between 74% 88%, and 93% in studies [20] and [21] using a Hidden Markov Model (HMM) classifier with small samples. The researchers of [22] used support vector machine learning to develop a classifier for Mandarin tones (T1, high-level; T2, low-rising; and T4, high-falling) recorded in two different auditory conceptions. Between 60% and 77%, accuracy was within reach. Machine learning methods are usually used to categorize ABR tests. For example, [23] classifies ABR into three groups based on the consonant vowels/da/, /ba/, and /ga/. To extract time-frequency characteristics from wavelet packet coefficients, the local discriminant basis (LDB) approach was used. The feature spaces were then reduced using the random subset feature selection (RSFS) approach. Discriminant analysis (DA), naïve Bayes (NB), multiclass support vector machine (MSVM), and K-nearest neighbors (KNN) approaches are used for classification. MSVM attained a maximum accuracy of 97%.

A significant amount of published research has utilized an ABR test dataset based on signal processing, although the latest study to be published [24] used an image rather than a signal, in keeping with our advised method. After preprocessing, which included scaling and cropping, they turned the PDF files from the archived reports into images. They then used Image Processing (IP) methods to segment each wave as a distinct wave image and transform all wave images into waves. The latency of the peaks for each wave was determined, and an audiologist may utilize this information to determine the condition.

Other studies used other data sets and stimuli to test DL-based classifiers. [16] This is the first attempt to categorize paired ABR waveforms without human narration and with little data using a deep convolutional neural network. It effectively achieved 92.9% accuracy, 92.9% sensitivity, and 96.4 % specificity.

### 1.3. Contributions

The novelty of this study is to compare the efficiency of numerous ML and DL algorithms to that of some tested detection methods using simulated data for which the ABR ground truth is known. Figure 1 shows the proposed framework in considerable detail. The following is a list of the contributions made using the suggested method:

1. In the study with challenging data collected, two different classification classes were performed, normal and abnormal.
2. Extensive research was conducted using AlexNet, DenseNet, Google Net, InceptionV2, InceptionV3, MobileNetV2, NASNetMobile, ResNet18, ResNet50, ResNet101, ShuffleNet, and SqueezeNet.
3. High accuracy rates were obtained.
4. With a small set of data, the TL was done instead of classical deep learning.
5. Accelerating the system and decreasing human error during the ABR test diagnosis.

The dataset used and the architectures in the classification study performed are described in section 2. Section 3 provides full experimental results. Finally, the result section is given in section 4.

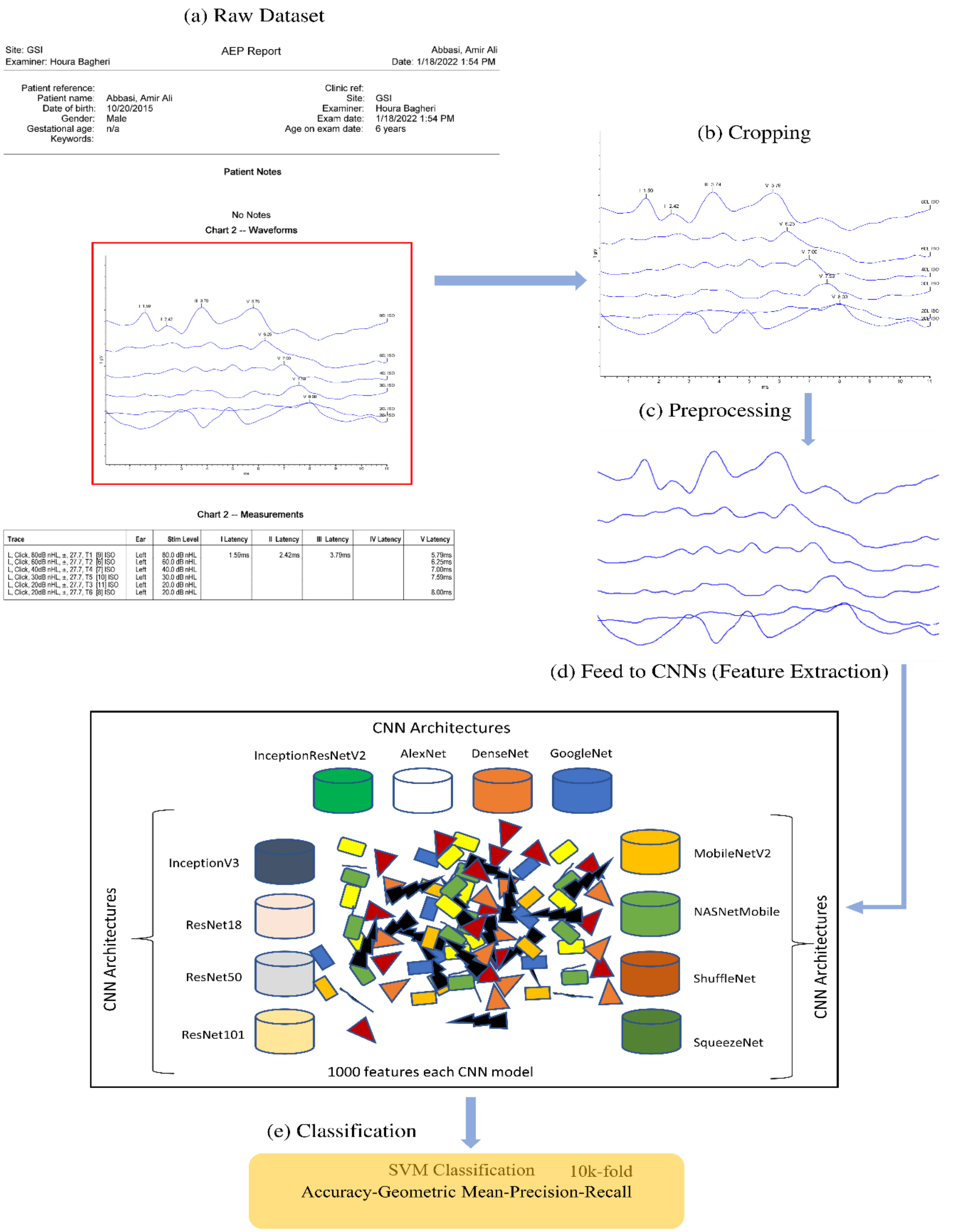

**Figure 1.** Proposed structure, a) reports on raw datasets stored in PDF format; b) extracting the relevant section of reports, including ABR waves; c) all unnecessary components, such as assessed peaks and axes, were removed during the preprocessing stage; d) all processed ABR images are fed into a set of 12 pre-trained deep learning models, which then select the most informative feature for each image based on TL; e) a normal/abnormal diagnostic using a 10k-fold SVM classifier.

## Material and Methodology

### 1.4. Dataset

The newborns varied in age from 1 month to 20 months (average age 4.34 months, standard deviation 4.01 months) (Out of the 187 image samples, 116 had normal hearing). The auditory evoked potential (AEP) was recorded as part of the neuropsychological battery utilizing Granson-Stadler (GSI) Audera brand equipment. We delivered a 100-second alternate click stimulation at 80 dB nHL and 27.71 clicks per second. The ER-3A (Etymotic Research) earphone inserts delivered stimuli to both the right and left ears. To record the ABR, three surface electrodes were used: one at the mid-forehead (Fz), one at the contest ear mastoid (ground), and one at the test ear mastoid as a reference. Each click reaction was averaged over 500-2000 repetitions, with two sets of data for each threshold level. The frequency range of the filter is 30 to 1500 Hz, and the time frame is 15 ms. In the initial phase of the study, two Audiology experts conducted independent reviews and marked the ABR peaks on the resulting waveform. The neurological labels normal and abnormal were determined, and the V-wave was traced to the cut-off point. Subsequently, the data were carefully compared, and we identified cases that exhibited similar results. In the next stage, when the two experts had diverging opinions, a third expert was consulted to thoroughly evaluate their perspectives and provide a conclusive opinion. Some of the aberrant waveforms had delayed activity peaks, tiny amplitudes, poor morphology and/or replicability, or worsened responses as the rate of sonic stimulation increased. In Figure 2, a few examples of the dataset are shown.

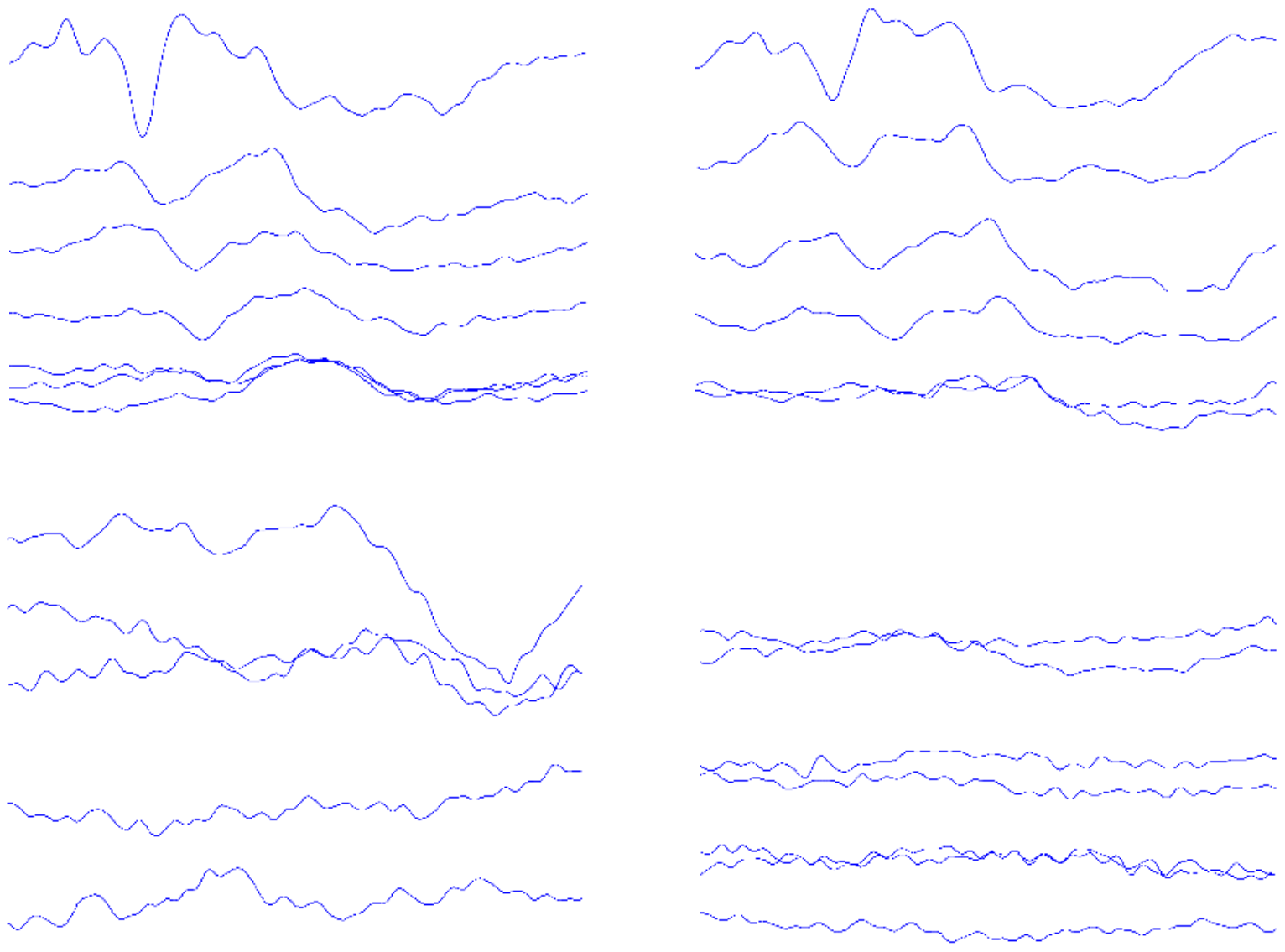

**Figure 1.** Samples of the dataset: first row normal images; second row abnormal images

## 1.5. Deep Learning Techniques

### 1.5.1. AlexNet Architecture

It was first designed by Hinton and his student Alex Krizhevsky in 2012, the winner of the ImageNet dataset, which consists of approximately 14 million image data from a thousand different classes [25]. It was observed that it was quite superior to traditional machine learning classification algorithms with its initial accuracy rates. The basis of the success of AlexNet is the ReLU activation function. ReLU-based networks train faster than other networks.

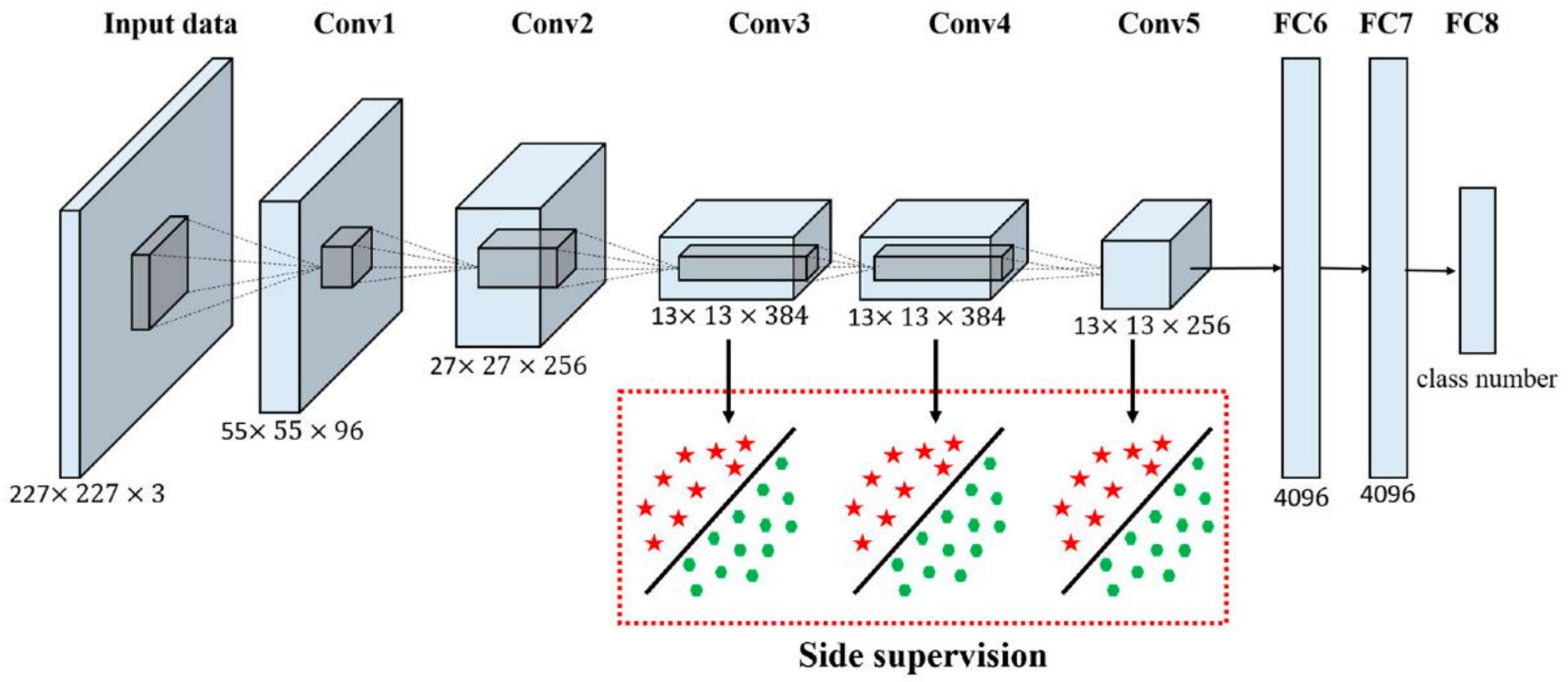

**Figure 3.** AlexNet Architecture [26]

AlexNet consists of 8 layers in total with learnable parameters (Figure 3). The model has 5 convolution layers with a combination of maximum pooling layers. After these convolutional layers, there are 3 fully connected layers. ReLU is used in all layers except the output layer. The activation function used in the output layer is Softmax. Dropout, on the other hand, is implemented in the first two fully connected layers.

The most important problem of AlexNet architecture is that it has 60 million parameters and this causes overfitting. To prevent this, the recommended methods are data augmentation and dropout.

AlexNet is one of the most important models of CNN for image classification, and it is one of the key pros of having images entered directly into the classification. But AlexNet isn't enough compared to the models that came after him.

### 1.5.2. DenseNet Architecture

DenseNet, a joint work of Cornwell University, Tsinghua University, and Facebook AI Research, is referred to in the literature as Dense Convolutional Network. The DenseNet article, first published in 2017, received more than 2000 citations and received the best article award. DenseNet is seen as a logical extension of the ResNet architecture [27]. The difference between ResNet and DenseNet is that the outputs are concatenated. There are many different types such as DenseNet-121 [28], DenseNet-169, DenseNet-161 [29], and DenseNet-201 [30].

In DenseNet architecture, all terms are interconnected. Each layer receives input from all other previous layers. In addition, it also communicates its property maps from itself to the next layer. This makes the number of channels less. This architecture also has higher computation and memory efficiency. Bottleneck levels, called DenseNet-B, reduce the complexity and scale of the model. This is processed before the ReLU-1x1 CONV layer ReLU-3x3 layer. Figure 4 illustrates the structural differences between ResNet and DenseNet.

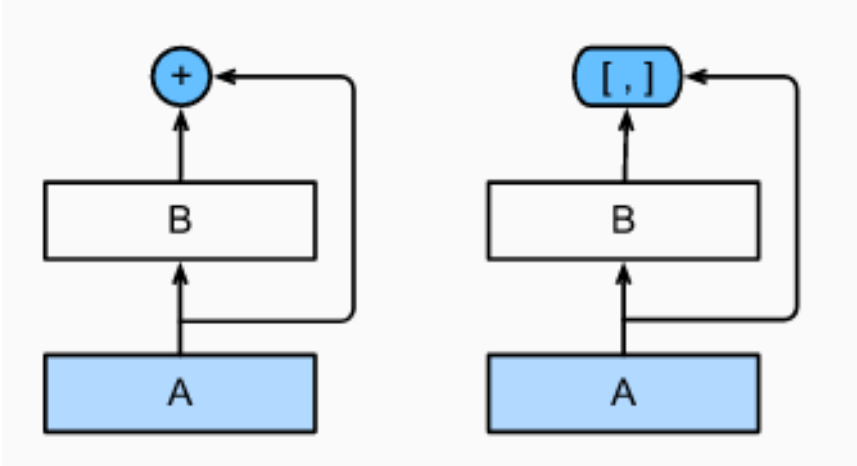

**Figure 4.** Difference between ResNet and DenseNet [31]

The DenseNet architecture has shown successful results even when the training data is insufficient [27]. This is because the classifier uses all of the complexity levels and thus tends to make clearer decision limits.

### 1.5.3. GoogleNet Architecture

GoogleNet architecture, which is mentioned in the literature as Parallel Concatenated Blocks, won the ImageNet competition held in 2014 [32]. This architecture has a complex structure consisting of Inception modules. Unlike other architectures, the depth and width of the network have been increased while the computational cost has been kept low.

The architecture consists of a total of 22 layers. To optimize quality, architectural decisions are based on the Hebbian principle and the intuition of multi-scale machining.

### 1.5.4. InceptionV3 Architecture

The InceptionV3 architecture is a convolutional neural network model. As in classical architecture, it consists of many convolution and pooling layers. In the last layer, there is a fully connected neural network. This architecture, which is suggested in the article “Rethinking the Inception Architecture for Computer Vision”, has 2 more models called InceptionV1 and InceptionV2 [33]. In this architecture, there is Batch normalization and a fully connected layer as an auxiliary classifier.

### 1.5.5. MobileNetV2 Architecture

MobileNet is one of the TL-based algorithms. There are three different models. The MobileNetV3 architecture, built based on the MobileNetV1 and MobileNetV2 architecture, is also available in two different versions, called MobileNetV3 Small and MobileNetV3 Large. It receives help from search optimization algorithms that belong to NAS and NetAdapt networks. The ReLU uses the h-swish activation function instead of the activation function [34]. The Swish function is an activation function such as the ReLU.

The MobileNet V2 architecture features linear bottleneck and inverted residual, which are used to extract attributes from high-size space without the loss of too much information. MobileNet V2 uses the bottleneck structure, a linear bottleneck layer, to reduce the input size.

### 1.5.6. ShuffleNet Architecture

Developed for mobile devices with very limited computing power, ShuffleNet is a CNN architecture that produces extremely successful results [35]. At the core of this architecture, accuracy is preserved. In addition, ShuffleNet uses pointwise group convolution and channel shuffle to reduce computational costs. Thus, feature map channels can encode more information and help reduce 1 x 1 convolution [36]. It achieves approximately 13 times more real speed than AlexNet, while at the same time providing similar accuracy.

### 1.5.7. SqueezeNet Architecture

The SqueezeNet architecture was presented in 2016 by Iandola et al., who worked as a researcher at DeepScale, University of California, Berkeley, and Stanford University in 2016 [37]. This architecture aims to create a neural network with fewer parameters and the architecture provides AlexNet-level accuracy with 50 times fewer parameters [38]. The SqueezeNet architecture has more efficient distributed layers and thanks to these layers, the workload in the neural network is reduced. The advantage of this situation is that it works faster [38, 39].

### 1.5.8. ResNet Architecture

ResNet was created as a result of a problem that researchers encountered when adding layers to the CNN architecture. As the layers are added to the CNN architecture, the performance increases up to a point, but decreases after a point. It is seen that gradient calculation is not performed in the ResNet model developed to solve this problem. Instead, a shortcut link is provided by arithmetically adding x to the f(x) function (Figure 5).

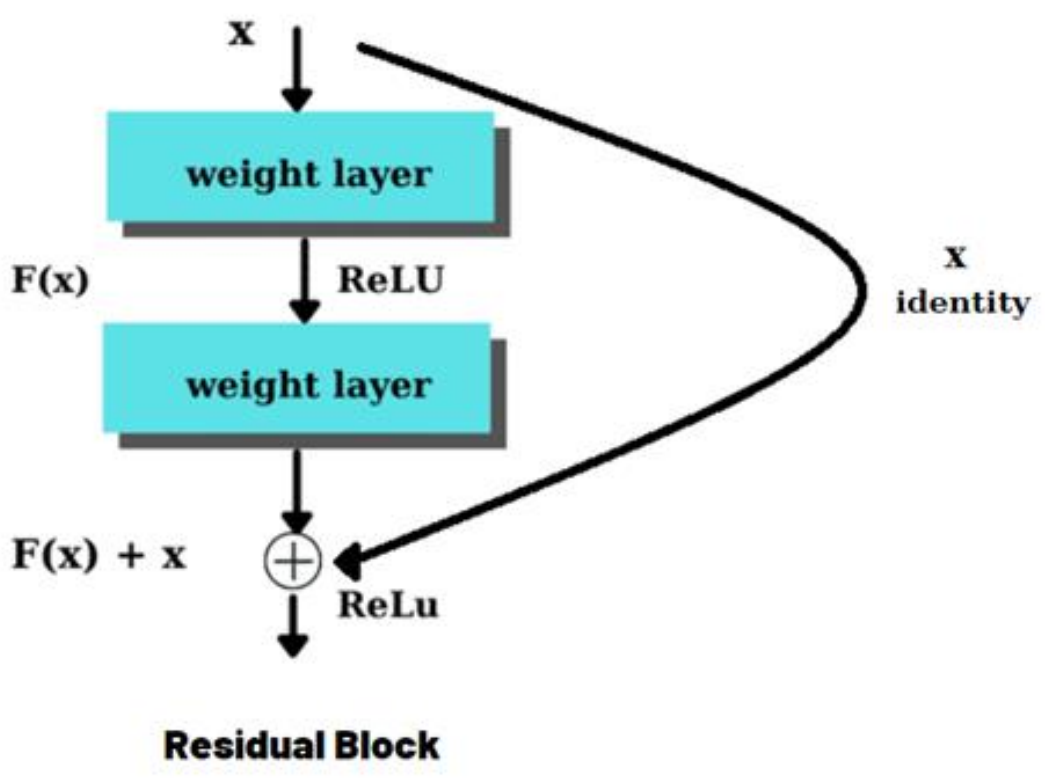

**Figure 5.** ResNet Architecture [40]

ResNet is a neural network architecture introduced by Kaiming He, Xiangyu Zhang, Shaoqing Ren, and Jian Sun in the article "Deep Residual Learning for Image Recognition" in 2015. It uses a 34-layer flat network architecture with fewer filters and lower complexity than VGG networks. Read More The architecture is then converted into a residual network by adding jump links or residual blocks to this flat network. There are many variants such as ResNet18, ResNet50, and ResNet101 used in the work done in this article [41].

Resnet 50 is obtained by replacing every 2-layer block in a 34-layer network with a 3-layer bottleneck block. These three layers have dimensions of 1×1, 3×3, and 1×1 [42].

A 101-layer ResNet is created using more 3-layer blocks. Now, with the help of networks, corruption problems are prevented. This resulted in significant accuracy gains from increased depth.

### 1.6. Evaluation Criteria

We used a variety of observational error metrics to assess how well each category worked. One metric used to evaluate classification models is accuracy. In this instance, "accuracy" refers to our model's adequately predicted rate. Precision is defined technically as follows:

$$Accuracy = \frac{TP+TN}{TN+TP+FN+FP} \quad (1)$$

where TP = True Positives, TN = True Negatives, FP = False Positives, and FN = False Negatives. Test results that accurately indicate the absence of a condition or trait are designated as TN, whereas those that accurately indicate the presence of a condition or feature are designated as TP. A FP test result shows that a certain condition or attribute is present, and a FN test result indicates that a certain condition or attribute is absent.

Precision and recall can be used to assess the utility of predictions. While recall measures the proportion of relevant items returned, precision measures the accuracy with which information is retrieved. Equations (2) and (3) define precision and recall (3).

$$Precision = \frac{TP}{TP+FP} \quad (2)$$

$$Recall = \frac{TP}{TP+FN} \quad (3)$$

The F-score (or F1-score) is a model performance metric that assesses how well a model performs on a specific dataset. The F-score combines a model's accuracy and recall and is defined as the harmonic mean of these two metrics. The equation describes the F1-score (4).

$$F1 - Score = \frac{2TP}{2TP+FP+FN} \quad (4)$$

The Geometric Mean (G-Mean) is a measure that measures the balance between classification performance in both more and fewer classes. The G-Mean is defined by the equation (5).

$$\mathrm{G-Mean} = \sqrt{\frac{TP * TN}{(TP + FN)(TN * FP)}} \quad (5)$$

The standard deviation (Std) relates to how much of the data is close to the mean. When the standard deviation is small, the data is close to the mean. The small standard deviation values mean that the obtained results are close to each other. The formula characterizes the Std (6), where $w$ is the number of data points, $x_i$ is each data point's value, and $\bar{X}$ is the mean of $x_i$

$$Std = \sqrt{\frac{\sum_{i=1}^{w}(x_i - \bar{X})}{w - 1}} \quad (6)$$

### 1.7. Transfer Learning

While it is true that a large dataset is required for training CNNs to achieve the desired accuracy, there are situations in which the difficulties associated with setting up a large dataset might reduce the model's performance accuracy. Obtaining matching training and test data in the actual world is notoriously challenging. The concept of "transfer learning" was proposed to help with this problem. This method is the most well-known machine learning method. The background pattern utilized to address the issue is employed in TL. The process involves first training the basic network on a relevant data set for a particular task, after which the learned weights are transferred to the desired network. Pre-trained model selection, problem size, and similarity are the two primary components of the TL process. Based on the issue associated with the target problem, the pre-trained model is chosen. The likelihood of overfitting is considerable if the target dataset is smaller (i.e., fewer than 1000 images) and comparable to the source training dataset (e.g., a dataset about biometrics, a vehicle dataset, a dataset of handwritten characters, etc.). The chance of overfitting is also minimal if the target data set is bigger and more comparable to the source data set, in which case the pre-trained model merely has to be fine-tuned [43, 44].

Twelve different CNN architectures are utilized to compare and contrast the features extracted via TL and fine-tuning. These twelve CNN architectures use the ImageNet data set as training material and use TL. This allows the architecture to pick up the generic features from various data sets with no further training. The normal and abnormal ABR are classified by a support vector machine using a set of features extracted separately from each CNN architecture.

## 2. Result and Discussion

In this piece, we use image data from ABR reports to propose a diagnostic model for the ABR test. As a result, the purpose of this study is to use the SVM classifier to compare the relative merits of feature extractions made by various deep learning techniques. To address this problem, we have developed the model shown in Figure 1.

This section describes the preprocessing, training, and testing processes. The experimental findings from the suggested procedures are then presented and discussed.

### 2.1. Preprocessing

The ear patient's response was recorded on a sheet using an Audera device as waves at various decibel (dB) levels. The device uses separate wave colours for a patient's right and left ears, red for the right, and blue for the left. The device's output report for each patient is saved as a PDF document file for audiology diagnosis. We gathered all of the PDF documents in this study and converted them to JPEG image files. To prepare the left and right ear images for each report for each patient in the dataset, the crop function as shown in Figure 1 is employed. Each cropped image was kept at a resolution of 2350 x 1950 pixels (width x height).

### 2.2. Training and testing phases

To get optimal outcomes, we investigated twelve pre-trained, highly effective deep neural network techniques. An imbalanced dataset from both the right and left ears was utilized to develop and evaluate the model. The dataset had 187 image samples (116 normal samples and 71 abnormal samples). Training and testing stages were implemented independently in two scenarios involving image feature extraction and diagnosing systems. All pre-trained models, including AlexNet, DenseNet, GoogleNet, InceptionV2, InceptionV3, MobileNetV2, NASNetMobile, ResNet18, ResNet50, ResNet101, ShuffleNet, and SqueezeNet, were utilized for feature extraction in the first scenarios. In the second scenario, the SVM classifier used all of the extracted features to diagnose normal and abnormal conditions. This research has used several different methods; however, the SVM classification produced the best results. Fivefold cross-validation (CV) was used throughout the whole paper to illustrate the effectiveness of our system. To achieve this, we trained and tested our system five times with different sets of data, utilizing 80% of the data for training and 20% for testing. We then determined the average of each fivefold CV. All classification outcomes are shown in Tables 1, 2, and 3.

### 2.3. Results

Experiments on the aforementioned dataset using the suggested models' SVM classifiers are summarized in Tables 1, 2, and 3. All achieved results are more than 90% system performance accuracy, demonstrating the excellent accuracy of all models. Throughout all experiments, the maximum accuracy and maximum GM success rate obtained is 95%. With the use of the ResNet50 and ShuffleNet models, this accuracy was attained. The ResNet50 model also achieved an average accuracy rate.

The lowest result in the work carried out was taken using the InceptionResNetV2 model. The max accuracy ratio obtained is 90%. The lowest mean accuracy ratio is 88%, also taken using the InceptionResNetV2 model.

**Table 1.** Maximum results of the proposed ABR classification in experiments using the imbalance dataset for all pre-trained models

| Models | Max Accuracy (%) | Max GM (%) | Max Precision (%) | Max Recall (%) |
|---|---|---|---|---|
| AlexNet | 94 | 93 | 94 | 93 |
| DenseNet | 94 | 93 | 94 | 93 |
| GoogleNet | 91 | 89 | 93 | 90 |
| InceptionResNetV2 | 90 | 88 | 91 | 89 |
| InceptionV3 | 94 | 92 | 95 | 92 |
| MobileNetV2 | 94 | 93 | 95 | 94 |
| NASNetMobile | 94 | 93 | 95 | 93 |
| ResNet18 | 94 | 93 | 95 | 93 |
| ResNet50 | 95 | 94 | 95 | 94 |
| ResNet101 | 93 | 91 | 93 | 91 |
| ShuffleNet | 95 | 94 | 96 | 94 |
| SqueezeNet | 93 | 91 | 93 | 92 |

**Table 2.** Average results of the proposed ABR classification in experiments using the imbalance dataset for all pre-trained models

| Models | Mean Accuracy (%) | Mean GM (%) | Mean Precision (%) | Mean Recall (%) |
|---|---|---|---|---|
| AlexNet | 89 | 87 | 89 | 88 |
| DenseNet | 91 | 91 | 92 | 91 |
| GoogleNet | 90 | 87 | 91 | 87 |
| InceptionResNetV2 | 88 | 85 | 88 | 86 |
| InceptionV3 | 91 | 89 | 93 | 90 |
| MobileNetV2 | 93 | 92 | 93 | 92 |
| NASNetMobile | 92 | 90 | 93 | 90 |
| ResNet18 | 92 | 91 | 92 | 91 |
| ResNet50 | 94 | 94 | 94 | 94 |
| ResNet101 | 90 | 88 | 90 | 88 |
| ShuffleNet | 92 | 90 | 93 | 91 |
| SqueezeNet | 91 | 90 | 91 | 90 |

**Table 3.** Std results of the proposed ABR classification in experiments using the imbalance dataset for all pre-trained models

| Models | Std Accuracy | Std GM | Std Precision | Std Recall |
|---|---|---|---|---|
| AlexNet | 0.0143 | 0.0167 | 0.016 | 0.016 |
| DenseNet | 0.0094 | 0.0118 | 0.0093 | 0.0113 |
| GoogleNet | 0.0079 | 0.0096 | 0.009 | 0.009 |
| InceptionResNetV2 | 0.0118 | 0.0146 | 0.0129 | 0.0136 |
| InceptionV3 | 0.008 | 0.0101 | 0.0089 | 0.0094 |
| MobileNetV2 | 0.0067 | 0.0073 | 0.0079 | 0.0071 |
| NASNetMobile | 0.0104 | 0.0127 | 0.0113 | 0.012 |
| ResNet18 | 0.0108 | 0.0130 | 0.0113 | 0.0124 |
| ResNet50 | 0.0050 | 0.0054 | 0.0058 | 0.0053 |
| ResNet101 | 0.0109 | 0.0132 | 0.0116 | 0.0126 |
| ShuffleNet | 0.0113 | 0.0133 | 0.0124 | 0.0128 |
| SqueezeNet | 0.0075 | 0.0084 | 0.0083 | 0.0082 |

As it can be seen in Tables 1, 2, and 3, experimental studies on 12 pre-trained CNN architectures were carried out. In each experiment, 5-fold cross-validation was used, and accuracy, geometric mean, precision, and recall parameters were calculated. Maximum, minimum, mean, and standard deviation values of these parameters were obtained by running each classifier 100 times. When the results obtained in the experimental studies are examined, it is seen that the accuracy of the ShuffleNet and ResNet50 architectures is at the highest rate, with 95% values. After these accuracy values, AlexNet, DenseNet, InceptionV3, MobileNetV2, NASNetMobile, and ResNet18 of the CNN architecture have the highest accuracy value of 94%. The standard deviation is less than 0.02% for the accuracy, geometric mean, precision, and recall values. The small standard deviation indicates that the results obtained with this classifier are close to each other.

We compared our suggested model with several studies that have the closest attributes to our dataset, such as stimuli and procedures, as indicated in Table 4, as we do not have access to the other dataset since the majority of published research employed a private dataset. By concentrating on the imbalance and the vast quantity of data, we were able to raise the model's accuracy to 95%.

**Table 4.** Comparison of our proposed model with the existing ABR classification frameworks

| Method | Sample size | Stimuli | Classifier | ABR feature | Accuracy (%) |
|---|---|---|---|---|---|
| Method proposed in [2] | 8 | 1 kHz and 4 kHz tone pips with a 2-cycle rise/fall time | CNN | Wave V of the transient response | 92.9 |
| Method proposed in [6] | 136 | click | Xgboost | features in the time-frequency | 92.0 |
| Method proposed in [17] | 130 | click | SVM | Discrete Wavelet Transform (DWT) | 97.0 |
| Method proposed in [45] | 83 | click | Pattern based | ABR time domain series | 97.6 |
| **proposed** | **187** | **click** | **SVM** | ShuffleNet | **95** |
| **proposed** | **187** | **click** | **SVM** | ResNet50 | **95** |

Without using any training epochs, all presented models converge in their underlying structures during feature extraction. These architectures make it possible to use CNN models for the most accurate feature extraction from ABR images. The suggested models for diagnosis of normal hearing thresholds and abnormal hearing (hearing loss) based on SVM classifier provide a high AUC value (higher than 90%), according to the values of false positive rate and true positive rate of the ROC curve displayed in Figure 6.

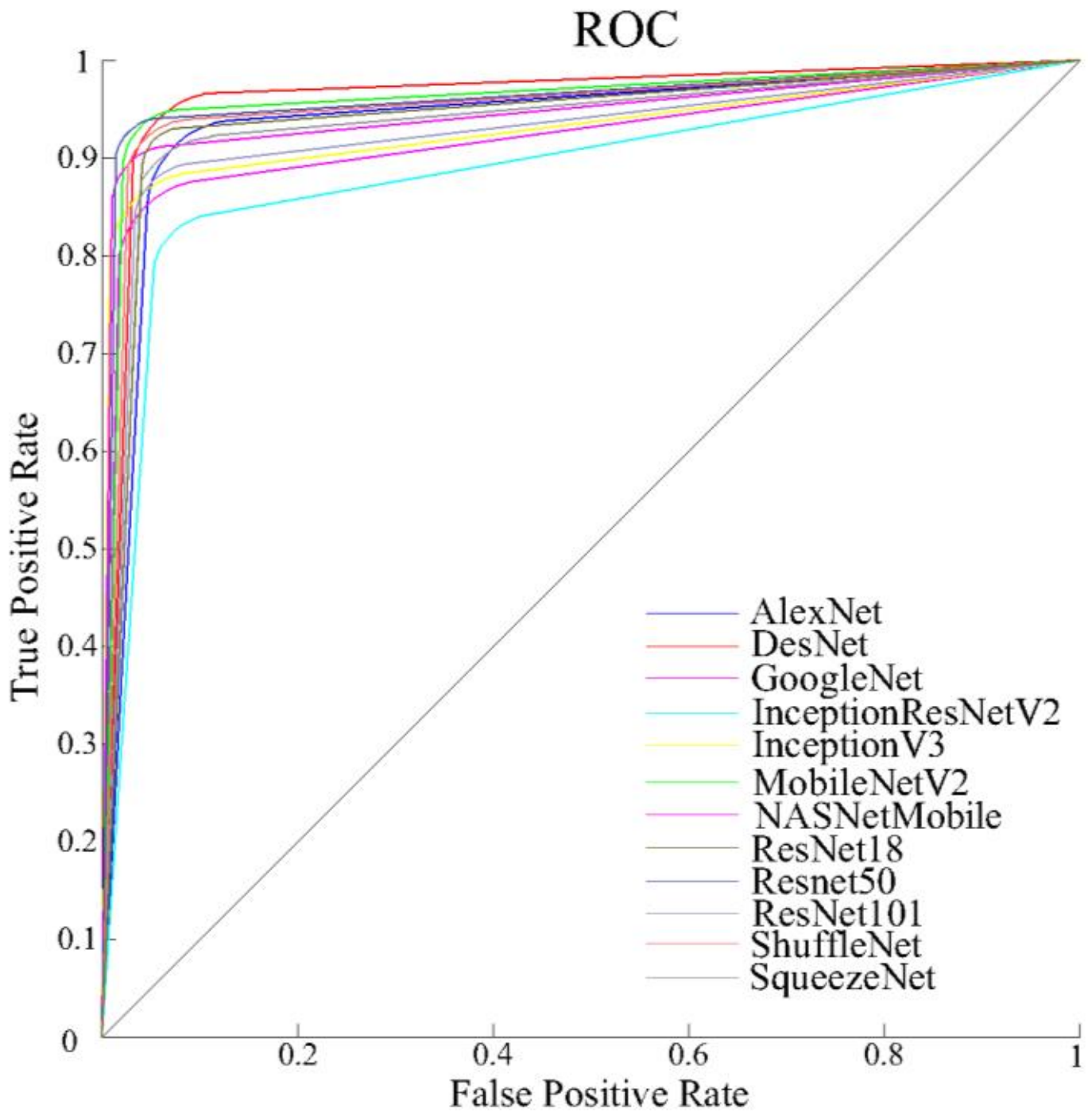

**Figure 6.** Estimating the ROC curve for the ABR feature extraction to improve models' performance

## Conclusion

The goal of this research is to collect a challenging dataset that uses ABR-derived PDF documents and images rather than waveforms on one side and to compare the efficiency of various ML and DL algorithms to that of some tested detection methods using simulated data for which the ABR ground truth is known on the other. The study with this especially collected data was done using TL instead of classical deep learning. A total of 12 models have been used and the success rate has generally been over 90%.

In the future, we'll work to address the scarcity of ABR data by utilizing GAN models that preserve the morphology and structure of ABR waves.

### Funding

There is no funding source for this article.

### Declaration of Interest

The authors declare that they have no known competing financial interests or personal relationships that could have appeared to influence the work reported in this paper.

### Credit Author Statement

Author[1]

Author[2]

Author[3]

Author[4]

## REFERENCES

[1] Hornickel, J., Chandrasekaran, B., Zecker, S., & Kraus, N. (2011). Auditory brainstem measures predict reading and speech-in-noise perception in school-aged children. Behavioural brain research, 216(2), 597-605.

[2] McKearney, R. M., & MacKinnon, R. C. (2019). Objective auditory brainstem response classification using machine learning. International Journal of Audiology, 58(4), 224-230.

[3] McKearney, R. M., Bell, S. L., Chesnaye, M. A., & Simpson, D. M. (2022). Auditory brainstem response detection using machine learning: a comparison with statistical detection methods. Ear and Hearing, 43(3), 949-960.

[4] Koravand, A., Al Osman, R., Rivest, V., & Poulin, C. (2017). Speech-evoked auditory brainstem responses in children with hearing loss. International Journal of Pediatric Otorhinolaryngology, 99, 24-29.

[5] Skoe, E., & Kraus, N. (2010). Auditory brainstem response to complex sounds: a tutorial. Ear and hearing, 31(3), 302.

[6] Wimalarathna, H., Ankmnal-Veeranna, S., Allan, C., Agrawal, S. K., Allen, P., Samarabandu, J., & Ladak, H. M. (2021). Comparison of machine learning models to classify Auditory Brainstem Responses recorded from children with Auditory Processing Disorder. Computer Methods and Programs in Biomedicine, 200, 105942.

[7] Ness, D. A. (2009). Normative data for neurodiagnostic Auditory Brainstem Response testing (ABR). Louisiana Tech University.

[8] Al Osman, R., & Al Osman, H. (2021). On the use of machine learning for classifying auditory brainstem responses: a scoping review. IEEE Access, 9, 110592-110600.

[9] Wimalarathna, H., Ankmnal-Veeranna, S., Allan, C., Agrawal, S. K., Samarabandu, J., Ladak, H. M., & Allen, P. (2022). Machine learning approaches used to analyze auditory evoked responses from the human auditory brainstem: A systematic review. Computer methods and programs in biomedicine, 107118.

[10] Huang, Y., Talwar, A., Chatterjee, S., & Aparasu, R. R. (2021). Application of machine learning in predicting hospital readmissions: a scoping review of the literature. BMC medical research methodology, 21(1), 1-14.

[11] Tuncer, T., Aydemir, E., Ozyurt, F., & Dogan, S. (2022). A deep feature warehouse and iterative MRMR based handwritten signature verification method. Multimedia Tools and Applications, 1-15.

[12] Özyurt, F., Ava, E., & Sert, E. (2020). UC-Merced image classification with CNN feature reduction using wavelet entropy optimized with genetic algorithm.

[13] Subasi, A., Mitra, A., Ozyurt, F., & Tuncer, T. (2021). Automated COVID-19 detection from CT images using deep learning. In Computer-Aided Design and Diagnosis Methods for Biomedical Applications (pp. 153-176). CRC Press.

[14] Chen, C., Zhan, L., Pan, X., Wang, Z., Guo, X., Qin, H., ... & Xiao, R. (2021). Automatic recognition of auditory brainstem response characteristic waveform based on bidirectional long short-term memory. Frontiers in Medicine, 7, 613708.

[15] Fallatah, A., & Dajani, H. R. (2018). Accurate detection of speech auditory brainstem responses using a spectral feature-based ANN method. Biomedical Signal Processing and Control, 44, 307-313.

[16] Valderrama, J. T., Alvarez, I., De La Torre, A., Carlos Segura, J., Sainz, M., & Luis Vargas, J. (2012). Recording of auditory brainstem response at high stimulation rates using randomized stimulation and averaging. The Journal of the Acoustical Society of America, 132(6), 3856-3865.

[17] Dobrowolski, A., Suchocki, M., Tomczykiewicz, K., & Majda-Zdancewicz, E. (2016). Classification of auditory brainstem response using wavelet decomposition and SVM network. Biocybernetics and Biomedical Engineering, 36(2), 427-436.

[18] Sadeghian, A., Dajani, H. R., & Chan, A. D. (2015). Classification of speech-evoked brainstem responses to English vowels. Speech Communication, 68, 69-84.

[19] Losorelli, S., Kaneshiro, B., Musacchia, G. A., Blevins, N. H., & Fitzgerald, M. B. (2020). Factors influencing classification of frequency following responses to speech and music stimuli. Hearing Research, 398, 108101.

[20] Llanos, F., Xie, Z., & Chandrasekaran, B. (2017). Hidden Markov modeling of frequency-following responses to Mandarin lexical tones. Journal of neuroscience methods, 291, 101-112.

[21] Llanos, F., Xie, Z., & Chandrasekaran, B. (2019). Biometric identification of listener identity from frequency following responses to speech. Journal of neural engineering, 16(5), 056004.

[22] Xie, Z., Reetzke, R., & Chandrasekaran, B. (2018). Taking attention away from the auditory modality: context-dependent effects on early sensory encoding of speech. Neuroscience, 384, 64-75.

[23] Shirzhiyan, Z., Shamsi, E., Jafarpisheh, A. S., & Jafari, A. H. (2019). Objective classification of auditory brainstem responses to consonant-vowel syllables using local discriminant bases. Speech Communication, 114, 36-48.

[24] Majidpour, A., Jameel, S. K., Majidpour, J., Bagheri, H., Rashid, T. A., Nazeri, A., & Aleaba, M. M. (2023). Detection of auditory brainstem response peaks using image processing techniques in infants with normal hearing sensitivity. Biomedical Signal Processing and Control, 86, 105117.

[25] Krizhevsky, A., Sutskever, I., & Hinton, G. E. (1970, January 1). ImageNet classification with deep convolutional Neural Networks. Advances in Neural Information Processing Systems. Retrieved December 6, 2022.

[26] Han, X., Zhong, Y., Cao, L., & Zhang, L. (2017). Pre-trained alexnet architecture with pyramid pooling and supervision for high spatial resolution remote sensing image scene classification. Remote Sensing, 9(8), 848.

[27] Huang, G., Liu, Z., Van Der Maaten, L., & Weinberger, K. Q. (2017). Densely connected convolutional networks. In Proceedings of the IEEE conference on computer vision and pattern recognition (pp. 4700-4708).

[28] Mao, W. L., Chen, W. C., Wang, C. T., & Lin, Y. H. (2021). Recycling waste classification using optimized convolutional neural network. Resources, Conservation and Recycling, 164, 105132.

[29] Theivaprakasham, H., Darshana, S., Ravi, V., Sowmya, V., Gopalakrishnan, E. A., & Soman, K. P. (2022). Odonata identification using Customized Convolutional Neural Networks. Expert Systems with Applications, 206, 117688.

[30] Wang, C., Sun, M., Liu, L., Zhu, W., Liu, P., & Li, X. (2022). A high-accuracy genotype classification approach using time series imagery. Biosystems Engineering, 220, 172-180.

[31] Li, Z., Lin, Y., Elofsson, A., & Yao, Y. (2020). Protein contact map prediction based on ResNet and DenseNet. BioMed research international, 2020.

[32] Simonyan, K., & Zisserman, A. (2014). Very deep convolutional networks for large-scale image recognition. arXiv preprint arXiv:1409.1556.

[33] Szegedy, C., Vanhoucke, V., Ioffe, S., Shlens, J., & Wojna, Z. (2016). Rethinking the inception architecture for computer vision. In Proceedings of the IEEE conference on computer vision and pattern recognition (pp. 2818-2826).

[34] Qian, S., Ning, C., & Hu, Y. (2021, March). MobileNetV3 for image classification. In 2021 IEEE 2nd International Conference on Big Data, Artificial Intelligence and Internet of Things Engineering (ICBAIE) (pp. 490-497). IEEE.

[35] Zhang, X., Zhou, X., Lin, M., & Sun, J. (2018). Shufflenet: An extremely efficient convolutional neural network for mobile devices. In Proceedings of the IEEE conference on computer vision and pattern recognition (pp. 6848-6856).

[36] Ajith, K., Menaka, R., & Kumar, S. S. (2021, May). EEG based mental state analysis. In Journal of Physics: Conference Series (Vol. 1911, No. 1, p. 012014). IOP Publishing.

[37] Iandola, F. N., Han, S., Moskewicz, M. W., Ashraf, K., Dally, W. J., & Keutzer, K. (2016). SqueezeNet: AlexNet-level accuracy with 50x fewer parameters and< 0.5 MB model size. arXiv preprint arXiv:1602.07360.

[38] Özyurt, F., Sert, E., & Avcı, D. (2020). An expert system for brain tumor detection: Fuzzy C-means with super resolution and convolutional neural network with extreme learning machine. Medical hypotheses, 134, 109433.

[39] Pathak, D., & El-Sharkawy, M. (2018, December). ReducedSqNet: A Shallow Architecture for CIFAR-10. In 2018 International Conference on Computational Science and Computational Intelligence (CSCI) (pp. 380-385). IEEE.

[40] Reddy, A. S. B., & Juliet, D. S. (2019, April). Transfer learning with ResNet-50 for malaria cell-image classification. In 2019 International Conference on Communication and Signal Processing (ICCSP) (pp. 0945-0949). IEEE.

[41] Liu, Y., She, G. R., & Chen, S. X. (2021). Magnetic resonance image diagnosis of femoral head necrosis based on ResNet18 network. Computer Methods and Programs in Biomedicine, 208, 106254.

[42] Ikechukwu, A. V., Murali, S., Deepu, R., & Shivamurthy, R. C. (2021). ResNet-50 vs VGG-19 vs training from scratch: A comparative analysis of the segmentation and classification of Pneumonia from chest X-ray images. Global Transitions Proceedings, 2(2), 375-381.

[43] Yang, L., Hanneke, S., & Carbonell, J. (2013). A theory of transfer learning with applications to active learning. Machine learning, 90, 161-189.

[44] Khan, S., Islam, N., Jan, Z., Din, I. U., & Rodrigues, J. J. C. (2019). A novel deep learning based framework for the detection and classification of breast cancer using transfer learning. Pattern Recognition Letters, 125, 1-6.

[45] Molina, M. E., Perez, A., & Valente, J. P. (2016). Classification of auditory brainstem responses through symbolic pattern discovery. Artificial intelligence in medicine, 70, 12-30.